\documentclass[sigconf]{acmart}
\usepackage{algorithm}
\usepackage{algpseudocode}
\usepackage[breakable]{tcolorbox}
\usepackage{tabularx}
\usepackage{multirow}
\AtBeginDocument{%
  }

\setcopyright{acmlicensed}
\copyrightyear{2025}
\acmYear{2025}
\acmDOI{XXXXXXX.XXXXXXX}

\acmConference[WWW 2025]{Make sure to enter the correct
  conference title from your rights confirmation emai}{April 28--May 02,
  2025}{Sydney, Australia}

\acmISBN{978-1-4503-XXXX-X/18/06}

\begin{document}


\title{A Large-scale Time-aware Agents Simulation for \\ Influencer Selection in Digital Advertising Campaigns}

\author{Xiaoqing Zhang}
\email{xiaoqingz@ruc.edu.cn}
\affiliation{%
  \institution{Gaoling School of Artificial Intelligence}
  \country{China}
}
\author{Xiuying Chen}
\email{xy-chen@pku.edu.cn}
\affiliation{%
  \institution{King Abdullah University of Science and Technology}
  \country{Saudi Arabia}
}
\author{Yuhan Liu}
\email{yuhan.liu@ruc.edu.cn}
\affiliation{%
  \institution{Gaoling School of Artificial Intelligence}
  \country{China}
}

\author{Jianzhou Wang}
\email{wangjianzhou@moonshot.cn}
\affiliation{%
  \institution{Moonshot AI}
  \country{China}
}

\author{Zhenxing Hu}
\email{huzhenxing@moonshot.cn}
\affiliation{%
  \institution{Moonshot AI}
  \country{China}
}

\author{Rui Yan}
\authornote{Corresponding authors.}
\email{rui.yan.pku@gmail.com}
\affiliation{%
  \institution{Gaoling School of Artificial Intelligence}
  \country{China}
}








\renewcommand{\shortauthors}{Trovato et al.}

\begin{abstract}
In the digital world, influencers are pivotal as opinion leaders, shaping the views and choices of their influencees. 
Modern advertising often follows this trend, where marketers choose appropriate influencers for product endorsements, based on thorough market analysis.
Previous studies on influencer selection have typically relied on numerical representations of individual opinions and interactions, a method that simplifies the intricacies of social dynamics.
In this work, we first introduce a Time-aware Influencer Simulator (TIS), helping promoters identify and select the right influencers to market their products, based on LLM simulation.
Specifically, after an advertisement is posted, we introduce a time simulation dimension into the process, which is rarely explored in previous simulation environments. 
This time-based simulation predicts user activity by modeling the user timeline and content lifecycle. It filters out most inactive users and contents, aligns with real-world behavior, and significantly reduces the scope of the simulation.
Each active user is represented as an LLM-based agent, drawing from their interaction history to deduce their profile and interests.
These user agents will predict their behavior in response to influencer advertising. 
After a period, we will obtain the interaction network generated from the advertisement's dissemination. Subsequently, we develop a ranking metric aimed at identifying influencers who have the highest potential to drive product purchases, based on the derived interaction network.
To validate our approach, we conduct experiments on the public advertising campaign dataset SAGraph which encompasses social relationships, posts, and user interactions. 
The results show that our method outperforms traditional numerical feature-based approaches and methods using limited LLM agents.
Our research shows that simulating user timelines and content lifecycles over time simplifies scaling, allowing for large-scale agent simulations in social networks.
Additionally, LLM-based agents for social recommendations and advertising offer substantial benefits for decision-making in promotional campaigns.
\end{abstract}

\ccsdesc[500]{Information systems~Social networks}


\keywords{Multi-agents, Time-aware Modeling, Influencer Selection}


\maketitle

\begin{figure}[htb]
    \centering
\includegraphics[width=1.0\linewidth]{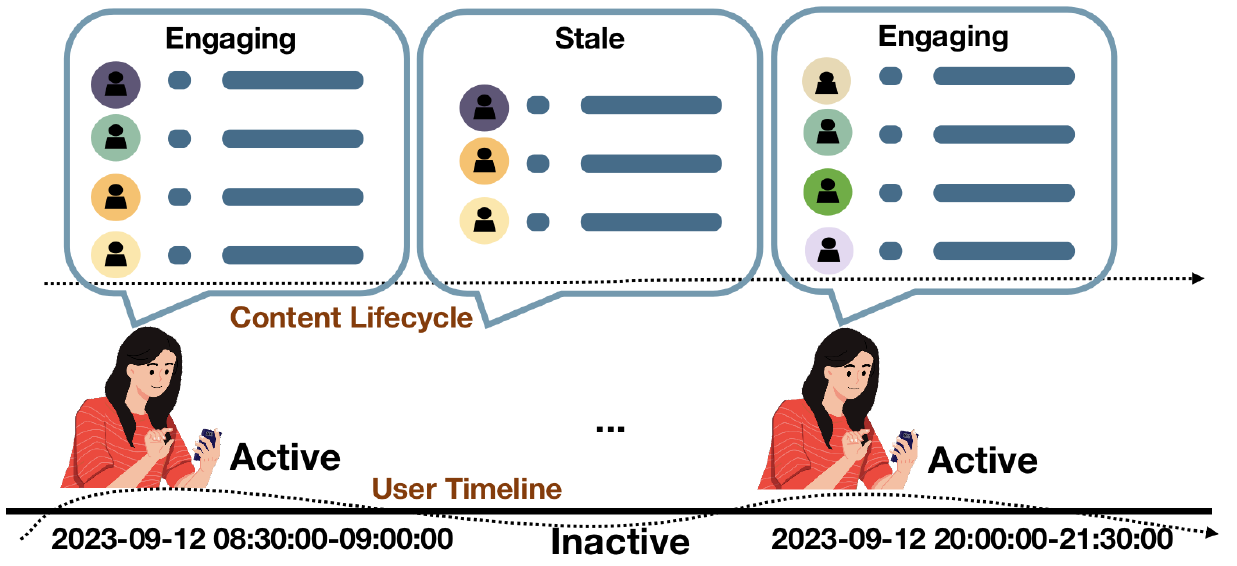} 
    \caption{The interaction-based user timeline and content lifecycle on the internet. ``Active'' refers to users interacting with the content they encounter. ``Inactive'' indicates that they are offline or disengaged from the content. ``Engaging'' content refers to content that users frequently interact with, whereas ``Stale'' content refers to outdated content that no longer attracts people to interact.}
    \label{fig:intro}
\end{figure}

\section{Introduction}
Social networks and social media platforms have revolutionized the way we interact and share information, creating vibrant online communities where influencers and their influencees engage in continuous dialogue. 
These influencers frequently make recommendations, whether for products, services, or lifestyle choices, which can have a considerable impact on their influencees' decisions and preferences. 
In this realm of digital advertising, promoters strategically select a limited number of influencers to maximize their product's influence and visibility in the market~\cite{zhang2015influenced,mallipeddi2022framework,lenger2022choose,qiu2018deepinf}.

Various methods have been developed for simulating opinions and selecting influencers in social networks. 
These approaches often reduce user opinions and interactions to numerical values and develop algorithms to predict changes in these numbers~\cite{zhu2021minimizing,zhou2023opinion,zhou2023opinion,dutta2020deep}. 
For instance, ~\cite{zhu2021minimizing} focuses on the optimization problem of suggesting new links to reduce polarization and disagreement in a social network, without altering initial opinions. 
However, these models often oversimplify reality, as people's opinions are diverse and nuanced, and the way others influence them is complex.
For example, individuals may be influenced by an influencer's perspective on one topic but not on another. 
Moreover, numerical representations fail to capture the nuances of textual dialogues, the reasons for opinion shifts, and the complexities of viewpoint evolution.~\cite{wang2022influential,chan2022study,gu2017co}.

Nowadays, researchers have been employing LLMs as agents for simulations~\cite{chen2023improving,chen2023topic,gao2024large,tornberg2023simulating}, and in this work, we utilize LLM simulation for the influencer selection task. 
Directly using an LLM for each user would result in tens of thousands of agents, leading to significant time and financial costs when deploying a large-scale agent society. 
Correspondingly, we note that although the network is large, the number of active users within a given time frame is limited. 
Figure \ref{fig:intro} shows an example of a user interaction timeline with social media content, illustrating that users are inactive from 9:00 AM to 8:00 PM. 
The content also has a lifecycle in that it spreads quickly when current and engaging but as engagement decreases and the content ages, it fades from the public.

With the above observations, in this work, we propose a Time-aware Influencer Simulator (TIS) for digital advertising campaigns, which aims to select suitable influencers for products by modeling LLM-based agents that track the temporal paths of both individuals and the content they generate.
The TIS provides several advantages compared with existing methods. 
Firstly, TIS effectively manages the number of active agents and content on social networks within a given time frame, reducing the simulation complexity and enabling the creation of a large-scale agent society.
Secondly, TIS constructs an interaction network for users at each time window, and as time progresses, this network continues to expand, accurately simulating the interactive behavior of users on social networks, which can be utilized for monitoring the progression of future events. 

Specifically, in TIS, we start by modeling the temporal trajectory, which includes modeling both the User Timeline (UT) and the Content Lifecycle (CL). 
UT is based on users' past interaction data, while CL relies on the interaction history of the content, i.e., the interaction time and counts towards the content.
By modeling the temporal trajectory, we can accurately estimate the number of active users needed for behavior simulation during specific periods and determine when the simulated content begins to lose its dissemination power.
Next, we defined the patterns for simulating user behavior with an LLM-based agent in the context of advertising recommendations. 
These patterns include self-awareness, social behavior prediction, and self-assessment. 
We anticipate that the agent will possess a clear understanding of itself, provide appropriate feedback on social discourse—either by commenting or by directly ignoring it—and accurately assess the level of support that simulated behaviors offer for the ultimate purchasing behavior related to the published advertisement.
Finally, in our simulation of influencer selection, we controlled the active agents based on the UT process and monitored their behaviors over a specified period. 
We focused on content identified by CL as being in its dissemination phase. 
By constructing a social network that evolves with time, we assessed the effectiveness of the advertisement's spread during the specified time frame, thereby reflecting the influence of the influencers.

We conducted experiments on the social advertising recommendation dataset SAGraph~\cite{zhang2024sagraphlargescaletextrichsocial}, which encompasses a wide range of user interactions and behaviors, providing a rich ground for analyzing and modeling the spread of influence and the effectiveness of advertising campaigns within various domains.
By leveraging this comprehensive dataset, we can delve into the intricacies of user engagement and the dynamics of social networks in a more realistic and nuanced manner. 

Our contribution can be summarized as follows: Firstly, we develop a temporal trajectory simulation mechanism, incorporating UT and CL modules, which provide a scalable solution for simulating large numbers of agents. Secondly, we propose a social network simulation framework driven by LLM agents that operate within specified time windows, enabling effective monitoring of information spread across social platforms. Finally, experiments in digital advertising campaigns demonstrate that our framework closely mirrors real-world influencer selection, reducing the need for extensive market surveys and aiding economic decision-makers in strategy development.

\begin{figure*}[tb]
    \centering
    \includegraphics[width=0.8\linewidth]{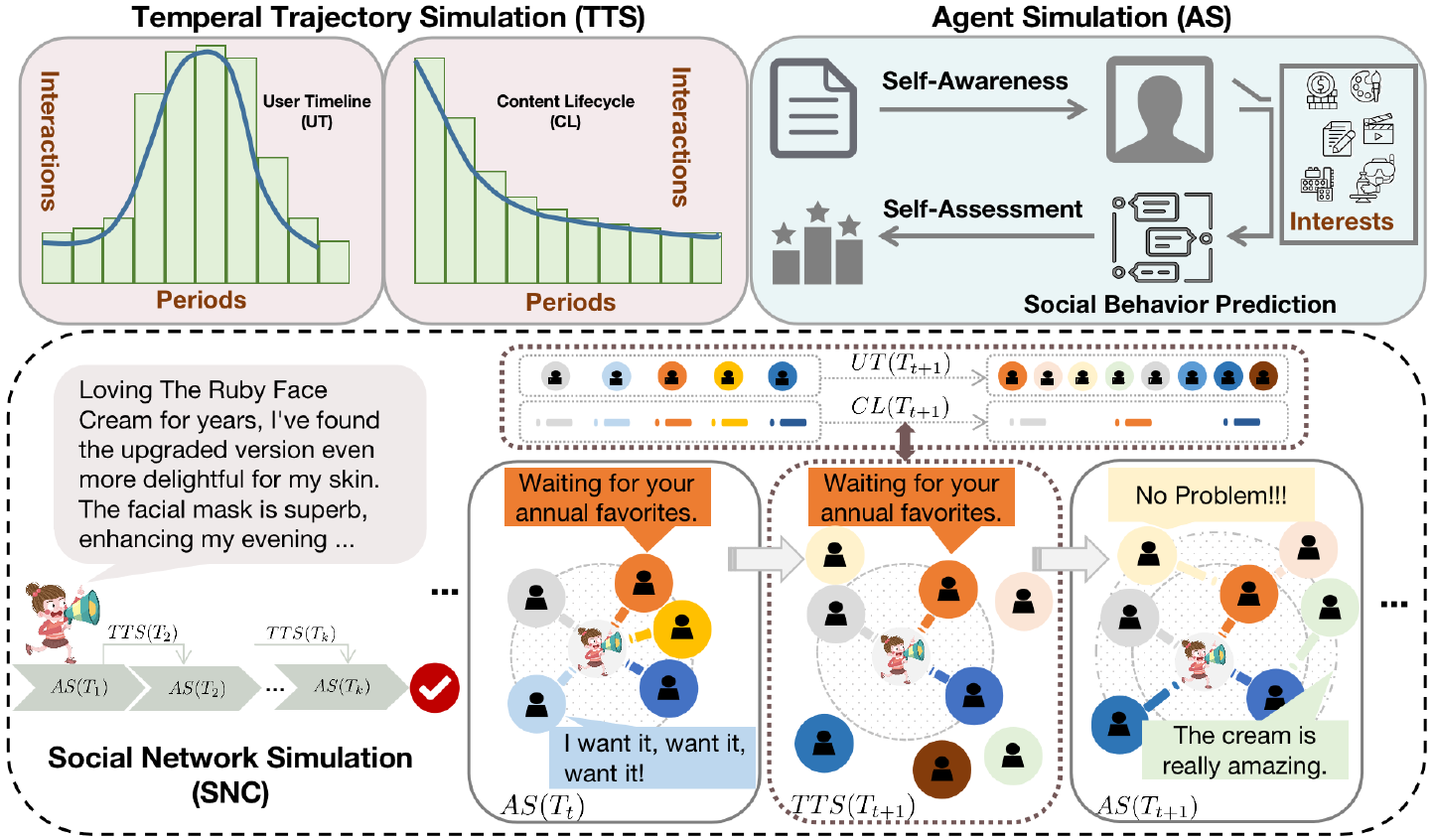}
    \caption{
    (a) Temporal Trajectory Simulation identifies agents and content to simulate within specific time frames, focusing on the user timeline and content lifecycle.  
    (b) Agent Simulation uses LLMs to simulate behaviors, including self-awareness, social behavior prediction, and self-assessment.  
    (c) Social Network Simulation leverages the TIS framework to model social network dynamics post-advertisement, offering insights into campaign effectiveness over time.}
    \label{fig:model}
\end{figure*}

\section{Related Work}

\textbf{Time-aware Modeling.} Time plays a crucial role in popularizing a topic and many studies on information dissemination in social networks rely on modeling time~\cite{iribarren2007information,lahiri2021todd,yang2010modeling}.
\citeauthor{abu2020time} introduces a time-aware framework that validates the credibility of users in a domain across different time dimensions through semantic and sentiment analysis of tweets, proving that it can be used to predict influential users within the domain.
\citeauthor{singh2020predicting} proposes a method based on random walks to predict information cascade, and it is verified that time is a key factor in social behavior.
\citeauthor{jia2022tt} utilizes the TT-Graph, which is based on text and time series, to model user activity trends in social networks.
However, the aforementioned time-based modeling only
stays at modeling events that have already occurred and cannot simulate unknown events at future times. 

\textbf{Multi-agents Simulation.} LLM-based multi-agents are employed to simulate social behavior, exploring social dynamics and information dissemination while reflecting real-world phenomena~\cite{park2023generative,gao2023s,chen2023multi,kaiya2023lyfe,liu2024skepticism,wang2024decoding,liu2024tiny}.
~\citeauthor{park2023generative} instantiate generative agents to create an interactive sandbox environment inspired by The Sims, allowing end users to engage with a small town of twenty-five agents using natural language. 
In addition, ~\citeauthor{park2022social} develop social simulation avatars, expanding the simulated community to 1,000 characters. 
Furthermore, ~\citeauthor{gao2023s} propose $S^{3}$ by constructing large networks with 8,563 and 17,945 agents, aimed at simulating social networks focused on gender discrimination and nuclear energy issues.
In contrast to their approach, we provide a more cost-effective solution for building larger-scale agent societies.

\textbf{User Influence Prediction.}
The task of selecting influencers within a social network is closely associated with predicting user impact on these platforms, wherein individuals who have the most significant influence and impact can be identified as potential influencers.
\cite{he2013improving} explores the correlation between web traffic volume and the number of tweets at different levels of detail.
They develop an optimization framework to extract traffic indicators from tweet semantics using a transformation matrix.
\cite{lampos2014predicting} devises a method to calculate a user's impact score by aggregating their followers, followees, and listings. 
Furthermore, \cite{rivadeneira2021predicting} develops an evidential reasoning prediction model that investigates how distinct features within Twitter posts can forecast the volume of retweets.
In our work, we have studied the changes in influencers' influence over time by simulating the construction of dynamic social networks, truly replicating the content dissemination paths in the real world.

\section{Methodology}
\subsection{Problem Formulation}
In our study, we simulated the social behaviors of tens of thousands of agents, studying how to select the best influencers for advertising campaigns in large social networks based on the purchase tendencies derived from the interaction network.
We simulate the interactions of followers of a key influencer $u$ on a social network following their exposure to an advertisement posted by $u$ over some time, and then observe the evolution of the interaction network that emerges from the advertisement. 
These interactions occur within a defined time horizon $T = [T_1, T_2, \ldots, T_k]$, where each $T_i$ represents a discrete time interval. 
Based on these interactions, we construct a dynamic interaction network $G_t = (V_t, E_t)$ in real-time, where $V_t$ denotes the set of participants engaging with the advertisement at time $t$ and $E_t$ represents the set of interactions among the participants within the network at that time.
The interactions $E_t$ are characterized by a collection of comments, represented by a series of tuples $(v, text, u)$, where $v$ is the users that participate in the interaction network with $u$ at time $t$ and $text$ is the comment associated with the user $v$.

To assess the influence of $u$, we calculate the purchasing inclination of all participants toward the product associated with the advertisement published by $u$, we utilize the behavioral and textual data from the interaction network. 
The individual purchasing inclinations across all participants after $T$ periods are measured by using a scoring function $S$, resulting in the total influence as $S(E_T)$.

By applying this rigorous framework, we can quantify the influence of the key influencer $u$ based on the collective purchasing behavior of their followers in response to the advertisement.
The overall process for the simulation of interaction towards the influencer $u$'s advertisement is shown in Algorithm~\ref{tis}.

\begin{algorithm}[tb]
\small
\caption{Time-aware Influencer Simulator (TIS)}
\label{alg:myalgorithm}
\begin{algorithmic}[1]

\State Let $u$ be the key influencer, $F_u$ be the set of followers, $H$ be the interaction history between $u$ and the followers
\State Generate profile $P_u$ for the influencer $u$
\State Generate profile $P_f$ for each follower in $F_u$
\State Set the periods to $T$
\State Let influencer $u$ post the $ad$ for product $P$
\State Initialize the interaction graph, set $\text{V=[[u]], E=[[(u, ad, None)]]}$
\For {each $t \in [1, ..., T]$}
    \State Get the active users by $UT(t)$
    \State Get the active content by $CL(t)$
    \State Set $V_t$ = $V_{t-1}$ + $UT(t)$
    \For {each active user $v$ in $V_t$}
        \State Retrieve profile $P_{v}$ of user $v$ from $P_f$
        \State Retrieve history $H_{v}$ between $v$ and $u$ from $H$
        \State Predict $v$' behavior reaction $R_t$ to $CL(t)$ based on $P_u$, $P_v$, $H_{v}$
        \State Set $E_t$ = $E_{t-1}$ + $E_{t}$
        \For {each reaction $(action, text)$ in $R_t$}
            \If{action is a comment}
                \State Add $v$ to $V_{t}$
                \State Add ($v, text, u$) to $E_t$
            \EndIf
        \EndFor
        \State Append $V_t$ to $V$
        \State Append $E_t$ to $E$      
    \EndFor
\EndFor
\State Calculate influence based on $E_T$ through the scoring function $S$
\end{algorithmic}
\label{tis}
\end{algorithm}

\subsection{Overall Architecture}
In Figure \ref{fig:model}, we summarize the overall architecture of our proposed Time-aware Influencer Simulator (TIS). It consists of three main components: the Temporal Trajectory Simulation (TTS), which includes the User Timeline module $UT(T_t)$ and the Content Lifecycle module $CL(T_t)$. This component focuses on identifying active users and content for each specific period $T_t$. The Agent Simulation (AS) utilizes user information to construct detailed profiles and simulates the interaction graph at period $T_t$ based on LLMs, defined as $AS(T_t)$. The Social Network Simulation (SNS) module utilizes $AS(T_t)$ during the $T_t$ period to construct the interaction network. It then proceeds to apply $TTS(T_{t+1})$ to update the network by incorporating active users and contents for the subsequent $T_{t+1}$ period.
This sequential execution ensures that the interaction network remains dynamic and reflective of the latest user engagement and content lifecycle.
We have each influencer publish advertisements and simulate interactions by $SNS$ over $T$ periods, resulting in the formation of the final interaction graph $G=(V_T, E_T)$. 
From this graph, we calculate the final influence using the scoring function $S(E_T)$. 
Ultimately, we select the top $k$ influencers based on these scores for comparison against actual promoters.

\subsection{User Timeline}
In modeling user timeline, we assume that the users' interaction behavior follows a certain time distribution, and we use the Gaussian Mixture Model (GMM)~\cite{reynolds2009gaussian} to model this distribution following \citeauthor{kirk2015towards,bohannon2020autoregressive,kalliovirta2015gaussian}.
The interaction times follow a certain distribution pattern, with peaks occurring in the early morning and evening which makes the GMM particularly suitable for capturing the trajectory characteristics of user interaction times~\cite{liu2019anomaly,raghavan2014modeling,salamzadeh2022grocery}. 
This section outlines the specifics of simulating daily user timelines based on interactions and modeling active users at any given time.

Suppose there are $N$ users, $\mathbf{x} = \{\mathbf{x}_i\}_{i=1}^{N} \in \mathbb{R}^{N \times T}$, where $\mathbf{x}_i = \{I_{1}, I_{2}, \ldots, I_{T}\} \in \mathbb{R}^{1 \times T}$ represents the interaction count vector for the $i$-th user over $T$ periods. Here, $I_t$ denotes the number of interactions of the $i$-th user during the $t$-th period.

We use a Gaussian Mixture Model (GMM) to model the distribution of user interactions over these periods. The assumption is that the interaction times come from a mixture of $K$ different Gaussian distributions. Each Gaussian component is parameterized by its mean $\mu_k$, variance $\Sigma_k$, and weight $\pi_k$, such that $\sum_{k=1}^{K} \pi_k = 1$. 

Here is the probability distribution of the user interaction data:
\begin{align}
    p(\mathbf{x}) = \sum_{k=1}^{K} \pi_k \cdot \mathcal{N}(\mathbf{x} \mid \mu_k, \Sigma_k)
\end{align}
where $\mathcal{N}(\mathbf{x} \mid \mu_k, \Sigma_k)$ denotes the $k$-th Gaussian distribution with mean $\mu_k$ and covariance matrix $\Sigma_k$, $\pi_k$ is the weight of the $k$-th Gaussian component.
The number of components $K$ corresponds to the number of peaks in user interaction time during the day.

To estimate the parameters $\Theta = \{\pi_k, \mu_k, \Sigma_k\}_{k=1}^{K}$, we apply the Expectation-Maximization (EM) algorithm on dataset $x$. 
The EM algorithm iteratively performs the expectation step (E-step) and the maximization step (M-step) to optimize the parameters until convergence, fitting the GMM to the distribution of user interaction times.
The probability density function $f(t)$ is derived from the GMM and represents the likelihood of user interactions at each period of the day. 
Given that the GMM has $K$ components, $f(t)$ is a weighted sum of these $K$ Gaussian distributions:
\begin{align}
    f(t) = \sum_{k=1}^{K} \pi_k \cdot \mathcal{N}(t \mid \mu_k, \Sigma_k)
\end{align}
where $\mathcal{N}(t \mid \mu_k, \Sigma_k) = \frac{1}{\sqrt{2\pi\Sigma_k}} e^{-\frac{(t - \mu_k)^2}{2\Sigma_k}}$ is the Gaussian function for component $k$ evaluated at period $t$, $\mu_k$ indicates the average time when users are most active for component $k$, $\Sigma_k$ reflects the variance or spread of user interactions around $\mu_k$, $\pi_k$ is the weight associated with each Gaussian component.





We use $f(t)$ to simulate the user timeline. The goal is to determine how many users interact during each period of the day and who these users are.
For each period $t$, we generate a random number $n_t$ according to a Gaussian distribution with parameters derived from $f(t)$. Firstly we compute the expected number of interactions for period $t$ as $N_t = \text{round}(N \cdot f(t))$, where $N$ is the total number of users. Secondly we randomly select $N_t$ user IDs from the total set of users to form the user ID set for that period, denoted as $UT(t)$:

\begin{align}
    UT(t) = \{ \text{ID}_1, \text{ID}_2, \ldots, \text{ID}_{N_t} \}
\end{align}

The user timeline process $UT(t)$ provides a collection $U = \{U_1, U_2, \ldots, U_{T}\}$, where $U_t$ is the set of interacting user IDs during the $t$-th period (with $t$ ranging from 1 to T). This process simulates user interactions throughout the day based on the fitted GMM, effectively capturing the peaks and patterns observed in the data.

\subsection{Content Lifecycle}
We propose the Content Lifecycle (CL) module based on the Cox Proportional Hazards Model (CoxPH)~\cite{bender2005generating} model to further reduce redundant data and speed up the simulation of content propagation. 
We define the set of unexpired content at time $t$ as $CL(t)$. 

Assume we have $N$ posts, denoted as $\mathbf{x} = \{\mathbf{x}_i\}_{i=1}^{N} \in \mathbb{R}^{N \times T}$. Each post $\mathbf{x}_i$ contains interaction counts recorded over $T$ periods, expressed as $\mathbf{x}_i = \{I_{1}, I_{2}, \ldots, I_{T}\} \in \mathbb{R}^{1 \times T}$. In this representation, $I_t$ denotes the number of interactions for the $i$-th post during the $t$-th period.
$X_i(t)$ represents the interaction history for post $i$ up to time $t$, which is the subset of $\mathbf{x}_i$ including all interactions recorded from the start until time $t$.
For each post $i$ based on its interaction data $\mathbf{x}_i$, we construct several lifecycle features:
\textbf{\textit{Interaction Start Time}} $t_{start, i}$ as the time when the $i$-th content first receives an interaction, \textbf{Duration} defined as the period from the content’s publication to when its interaction count first becomes zero, denoted as $d_i = \min\{t | X_{i,t} = 0\} - t_{start, i}$, \textbf{\textit{Average Interaction Count}} during the period when the content has interactions (from $t_{start, i}$ to $d_i$), the average interaction count is $\bar{X}_i = \frac{1}{d_i} \sum_{t=t_{start, i}}^{d_i} X_{i,t}$, \textbf{\textit{Total Interaction Count}} represents the total interaction count from the content’s publication to its expiration is $S_i = \sum_{t=t_{start, i}}^{d_i} X_{i,t}$, \textbf{\textit{Minimum Interaction Count in a Fixed Interval}} for a fixed time interval $\Delta$ (e.g., one minute), the minimum interaction count within the interval is $m_i = \min\left\{\sum_{j=0}^{\Delta-1} X_{i,t+j} \mid t_{start, i} \leq t \leq d_i - \Delta \right\}$.

These features serve as covariates $F = \{t_{start, i}, d_i, \bar{X}_i, S_i, m_i\}$ for the CoxPH model, describing the lifecycle characteristics of each content.
The CoxPH model estimates the risk of the content expiring at time $t$. 
Assuming the hazard function is $h(t|F)$, where $F$ represents the covariates, the Cox model is defined as:
\begin{align}
    h(t|F) = h_0(t) \exp(\beta_1 F_1 + \beta_2 X_2 + \ldots + \beta_p F_p),
\end{align}
where $h_0(t)$ is the baseline hazard function that captures the intrinsic risk of an event occurring over time when covariates are zero, $\beta_i$ is the coefficient corresponding to covariate $F_i$, reflecting the influence of that covariate on the expiration risk.


We input each content’s interaction data and features into the CoxPH model. 
First, we define the expiration time of the content as the moment when its interaction count is zero for multiple consecutive periods, i.e., $\tau_i = \min\{t | X_{i,t} = 0 \text{ and consecutive}\}$. 
We then use these events and features for model training, fitting the model’s coefficients $\beta$ by maximizing the log-likelihood function.

After training, we use the CoxPH model to predict the new content's survival time as follows:
\begin{align}
    \hat{P}(t | F) = 1 - \exp\left(-\int_0^t h_0(s) \exp(\hat{\beta}^T F) ds\right),
\end{align}
where $\hat{P}(t|F)$ is the probability that content expires before time $t$.


At any given time $t$, the set of unexpired content $CL(t)$ is formed based on the survival probabilities and expiration times:
\begin{align}
CL(t) = \{i \in \{1, 2, \ldots, N\} \mid \hat{P}(t | F(X_i(t))) > 0 \text{ and } t < \tau_i \},
\end{align}
where $X_i(t)$ represents the post $X_i$'s interaction history at time $t$.

\begin{table*}[t!]
    \centering
    \resizebox{0.8\textwidth}{!}{%
        \begin{tabular}{l cccccc | cccccc}
            \toprule
\multirow{3}{*}{Models} & \multicolumn{6}{c}{\textsf{Spark Thinking}} & \multicolumn{6}{c}{\textsf{Intelligent Floor Scrubber}}\\
\cmidrule(lr){2-7} \cmidrule(lr){8-13}
& P@5 & P@10 & R@5 & R@10 & G@5 & G@10 & P@5 & P@10 & R@5 & R@10 & G@5 & G@10\\
\midrule
TIS & \textbf{0.60} & \textbf{0.40} & \textbf{0.50} & 0.67 & \textbf{0.70} & \textbf{0.71} & \textbf{0.60} & \textbf{0.40} & \textbf{0.60} & \textbf{0.80} & \textbf{0.70} & \textbf{0.81}\\
GPT-4 w/ profile\&CoT & 0.40 & 0.30 & 0.50 & 0.75 & 0.56 & 0.68 & 0.40 & 0.30 & 0.50 & 0.75 & 0.59 & 0.72\\
CELF  & 0.20 & 0.20 & 0.25 & 0.50 & 0.15 & 0.28 & 0.00 & 0.20 & 0.00 & 0.50 & 0.00 & 0.26\\
CELF++  & 0.40 & 0.20 & 0.50 & 0.50 & 0.64 & 0.64 & 0.40 & 0.20 & 0.50 & 0.50 & 0.64 & 0.64\\
SIGMA  & 0.00 & 0.20 & 0.00 & 0.50 & 0.00 & 0.23 & 0.20 & 0.20 & 0.25 & 0.50 & 0.25 & 0.36\\
PI  & 0.00 & 0.20 & 0.00 & 0.50 & 0.00 & 0.23 & 0.20 & 0.20 & 0.25 & 0.50 & 0.25 & 0.36\\
             \midrule
\multicolumn{1}{c}{} & \multicolumn{6}{c}{\textsf{Ruby Face Cream}}& \multicolumn{6}{c}{\textsf{SUPOR Boosted Showerhead }} \\
    \cmidrule(lr){2-7}  \cmidrule(lr){8-13}
TIS     & \textbf{1.00} & \textbf{0.80} & \textbf{0.50} & \textbf{0.80} & \textbf{1.00} & \textbf{0.86}         & \textbf{0.40} & \textbf{0.30} & 0.33 & \textbf{0.50} & 0.38 & \textbf{0.45} \\
GPT-4 w/ profile\&CoT     & 0.60 & 0.50 & 0.43 & 0.71 & 0.53 & 0.60         & 0.40 & 0.20 & 0.50 & 0.50 & 0.40 & 0.40 \\
CELF    &0.40 & 0.40 & 0.29 & 0.57 & 0.35 & 0.46           & 0.20 & 0.20 & 0.25 & 0.50 & 0.20 & 0.33 \\
CELF++     & 0.40 & 0.20 & 0.29 & 0.29 & 0.51 & 0.41      & 0.20 & 0.10 & 0.25 & 0.25 & 0.25 & 0.25 \\
SIGMA    & 0.60 & 0.40 & 0.43 & 0.57 & 0.53 & 0.53         & 0.20 & 0.10 & 0.25 & 0.25 & 0.39 & 0.39 \\
PI       &0.60 & 0.40 & 0.43 & 0.57 & 0.66 & 0.63          & 0.20 & 0.10 & 0.25 & 0.25 & 0.39 & 0.39 \\
            \bottomrule
        \end{tabular}}

    \caption{Performance comparison of our model and baselines. The TIS framework provides results that contain statistically significantly
(p < 0.05) compared to the baselines on all products.}
    \label{tab:main}
\end{table*}

\subsection{Agent Simulation}
The behavior simulation of agents is divided into three stages: self-awareness, social behavior prediction, and self-assessment. Below is a detailed introduction to each component.

\noindent $\bullet$
\textbf{Self-Awareness.}
An ideal product advertiser should be an influencer who shares the same interest domain as the product.
This ensures that the influencer's influencees trust the endorsement. 
Otherwise, a mismatch between the influencer's focus and the product may lead to doubts among influencees about the influencer's expertise in that specific domain, potentially affecting the outcome.

In this regard, the LLM-based agent needs to generate personalized profiles through reasoning.
This approach is chosen as these attributes provide a comprehensive overview of the influencer, reflecting their long-term characteristics and potential influence~\cite{peng2018influence}.

\noindent $\bullet$
\textbf{Social Behavior Prediction.}
Once the user profile for each agent is obtained, these agents can proceed to simulate the behavior of each user in response to the posts that are available to them.
Predicting user behavior based on user profiles has been widely explored and verified~\cite{wall1991predicting,fabra2020log}.
For example, \cite{sharma2022role} collects a large-scale Tweet dataset to train a retweet prediction model. 
In contrast, our agent-based simulation eliminates the reliance on traditional training methods and provides textual prediction.

\noindent $\bullet$
\textbf{Self-Assessment} 
For the simulated behaviors of agents, we combine the objective of influence calculation to self-assess the simulated behaviors' impact on the purchase inclination towards the products involved in the advertisement. This is the feedback of agent simulation, achieving the quantification of agent behavior towards the ultimate goal. 
Ultimately, assessments from multiple periods are aggregated to determine the influence of the published advertisement on multiple agents by the influencers. 

\subsection{Social Network Simulation}
The TIS framework is a dynamic process of social network simulation (SNC) that simulates user behaviors and interactions within a social network to evaluate and predict changes in influence. 
This approach allows us to understand how information spreads and how user behavior and decision-making can be influenced. 
From the moment the influencer $u$ publishes the advertisement $ad$, the advertisement begins to disseminate through the network, and we observe this process over $T$ time periods.

At time $T_t \in T$, the active agents are simulated by $UT(T_t)$. 
They interact with each other and produce the comments to form the interaction network $AS(T_t)$.
Before the $t+1$ round of simulated interactions begin, it is necessary to perform real-time deduplication operations on the network. 
The UT process is used to update the active users at time $t+1$, while the CL strategy is used to update the active content at time $t+1$. 
We update $AS(T_t)$ with $UT(T_{t+1})$ and $CL(T_{t+1})$ to get the interaction network $TTS(T_{t+1})$ through TTS process. 
Then we start agent simulation for the $t+1$ round.
By iterating through $T$ periods in this manner, we obtain the interaction network $G_T = (V_T, E_T)$ resulting from the dissemination of the advertisement $ad$ published by the influencer $u$. 
Then we transform the behaviors $A = \{\text{action, text}\}$ generated at time t into influence for product promotion.
We only consider the behaviors when $action=comment$.

We define the number of interacting counts $S_N=|E_T|$, the purchase inclination $P_i$ of each comment in $E_T$, the average purchase inclination $\bar{P}$ is calculated as $\bar{P} = \frac{\sum_{i=1}^{S_N} P_i}{S_N}$.
While the standard deviation $\sigma$ measures the variability of these time-series inclinations from the average, calculated as $\sigma = \sqrt{\frac{\sum_{i=1}^{S_N} (P_i - \bar{P})^2}{S_N}}$.

The scoring system $S$ consists of two components: the interaction user count score $S_N$, representing the total number of users interacting, and the purchase inclination consistency score $ S_{\sigma}$, which measures the concentration of purchase inclinations. The consistency score is calculated based on the standard deviation, normalized against the maximum standard deviation observed across all interactions, as follows:
\begin{align}
    S_{\sigma} = 1 - \frac{\sigma}{\sigma_{\text{max}}}.
\end{align}

This ensures that a smaller standard deviation yields a higher score, reflecting a greater consistency in purchase inclinations.
We balance the interaction count and purchase inclination consistency with a weighted formula:
\begin{align}
    S = 
    \begin{cases} 
        S_N & \text{if } S_N < \theta \\
        \alpha \cdot S_N + (1 - \alpha) \cdot S_{\sigma} & \text{otherwise}
    \end{cases}
\end{align}
Here, $\alpha$ is a weight between 0 and 1, allowing for flexibility in emphasizing either the number of interactions or the consistency of purchase intentions based on the specific context or goals of the advertising campaign. $\theta$ represents the minimal interaction counts for considering the consistency of purchase inclination. This multi-faceted approach not only accounts for user engagement with the advertisement but also evaluates how reliably those users are inclined to make purchases, ultimately providing a more nuanced assessment of potential advertising promoters.

\section{Experiment}

\subsection{Dataset}
We simulate and evaluate our framework on SAGraph~\cite{zhang2024sagraphlargescaletextrichsocial}, which is a large-scale text-rich social graph dataset that aims to assess the complex process of influencer selection in advertising campaigns and highlights the significance of text in advertising. 
It includes products spanning various domains. 
Each domain encompasses 40 to 300 key influencers, each with follower counts approaching 100,000. 
Additionally, each domain involves hundreds of thousands of users. 
This multi-domain dataset includes real interaction networks and interaction texts, with interaction information collected over several days. 
This temporal span makes the dataset well-suited for modeling the time-aware social network's spreading.

\subsection{Baselines}
We compare TIS with the following baselines.

$\bullet$\textbf{GPT-4 w/ profile\&CoT}~\cite{zhang2024sagraphlargescaletextrichsocial} utilizes the limited LLMs as the influencer dynamics simulator, helping promoters identify and select the right influencers to market their products.

$\bullet$\textbf{CELF}~\cite{leskovec2007cost} is an optimization technique for influence maximization in social networks, enhancing greedy algorithm efficiency by reducing redundant computations.

$\bullet$\textbf{CELF++}~\cite{goyal2011celf++} further improves CELF's efficiency by exploiting submodularity.

$\bullet$\textbf{SIGMA}~\cite{yan2019minimizing} selects blocker nodes in rumor blocking, this model is adapted for our advertising scenario to minimize influence.

\subsection{Evaluation Metrics}
We adopt standard Top-$k$ ranking metrics for evaluation. These metrics include Precision ($P@k$), which measures the proportion of relevant items in the top-$k$ recommendations; Recall ($R@k$), which assesses how many of the total relevant items are included in the top-$k$ recommendations; and Normalized Discounted Cumulative Gain ($G@k$), which values correct recommendations more favorably based on their position in the ranking for evaluation.

\subsection{Implementation Details}
We set the number of components $K=2$ during the GMM modeling process, which was determined by the observed peaks in interaction distribution during the early morning and evening. 
During the CL process, we set $\tau_i=10$.
Each time interval $t$ is set to be one minute, and due to cost considerations, we continued to simulate a period of $T = 100$. 
We employ GPT-4 for agent simulation.
When simulating user behavior in the LLM, we adopted a method of individually simulating single users instead of batch-simulating multiple users to avoid interference between users. 
Considering the large base of interacting users in the influence calculation phase, we set $\alpha = 0.02$ and $\theta=15$ to balance the impact of the increasing number of interactions on influence calculation as time and the user grows.
\begin{figure}[tb]
    \centering
\includegraphics[width=1.0\linewidth]{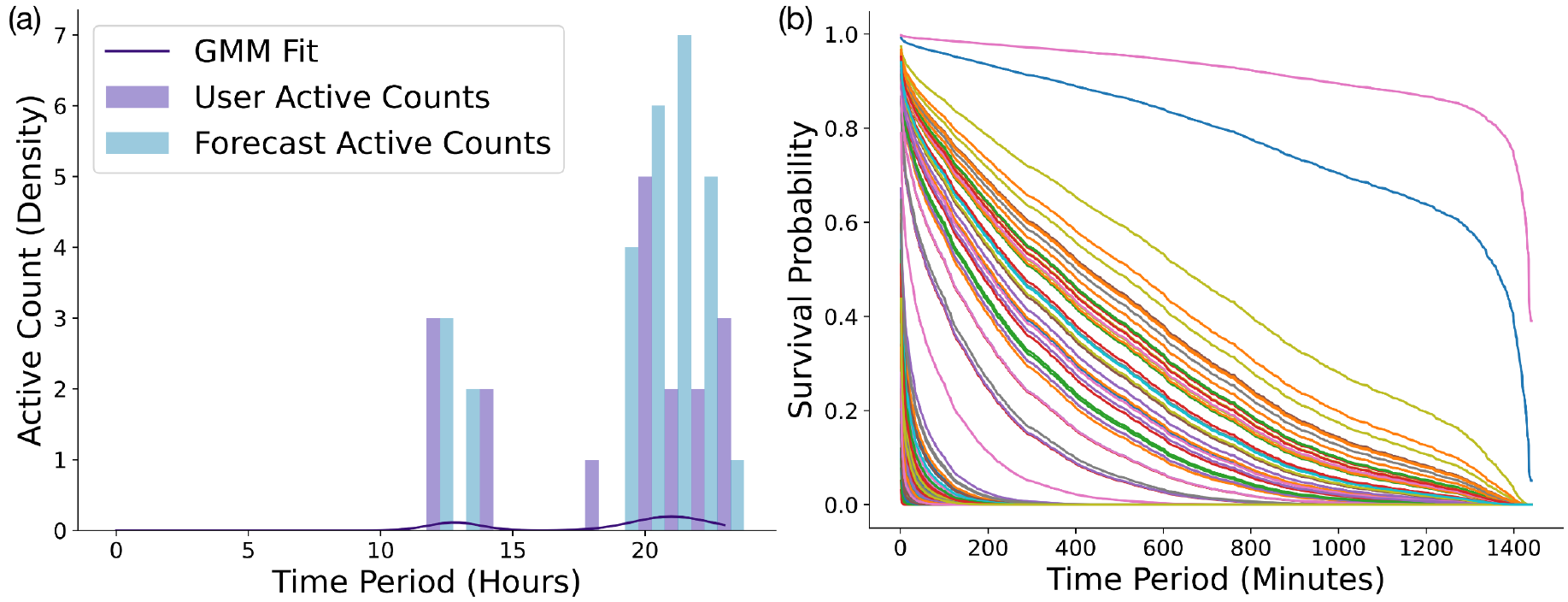} 
    \caption{(a) User timeline modeling. ``GMM Fit'' represents the simulated curve, ``User Active Counts'' indicates the actual interactions engaged in each period of the day, and ``Forecast Active Counts'' signifies the predicted number of interactions for a certain user. (b) Simulation of the content lifecycle, where each line represents the survival probability distribution of a post or comment over a day, which is 1,440 minutes.}
    \label{fig:utr}
\end{figure}

\begin{figure*}[htb]
    \centering
\includegraphics[width=0.85\linewidth]{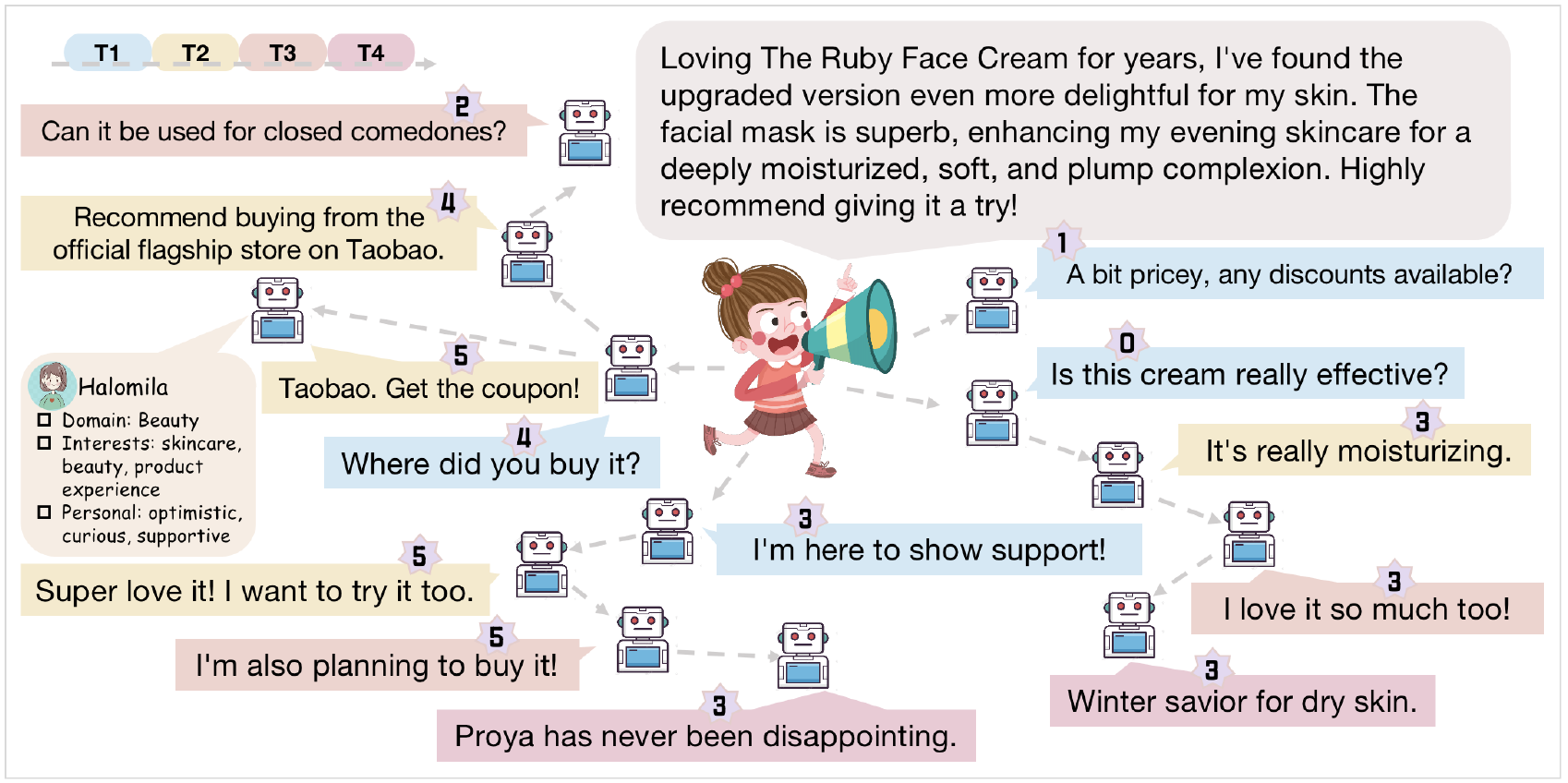} 
    \caption{The illustration of the interaction network after 4 periods’ simulation, starting with the influencer posting an advertisement for Ruby Face Cream. The colored dialogue boxes represent the agent's comments from different periods. The scores for the generated comments reflect the inclination to purchase, starting from no intention (0) to a definite plan to buy (5).}
    \label{fig:case}
\end{figure*}

\subsection{Overall Performance}
We show the overall performance of TIS and baselines in Table \ref{tab:main}.

In preliminary experiments, we find that baseline models performed inadequately, often failing to identify any positive influencers.
To enhance their capability, we adopted the pre-selection strategy proposed by ~\citeauthor{zhang2024sagraphlargescaletextrichsocial} into the baselines to consider only the top candidates from the Pre-selection as selected influencers. 
However, even with pre-selection, the baselines significantly underperform compared to our model. 
This shortfall could stem from the baselines' simulation logic, which emphasizes the number of nodes an influencer can reach and the extent of their subsequent network impact, without considering the semantic alignment between the influencer's content and the products.

Secondly, we can see that our model surpasses the strong baseline, GPT-4 w/ profile\&CoT, in the NDCG@5 and NDCG@10 metrics. 
Except for the SUPOR Boosted Showerhead, all products have shown an improvement of over 10 percentage points.
Compared to digital baselines like CELF, CELF++, SIGMA, and PI, one aspect of our advantage stems from the LLM's ability to analyze social text. 
In contrast, when compared to LLM-based baselines, the other aspect of our advantage arises from the strengths of our architecture, which allows for a more realistic simulation of content dissemination and the evolution of social networks.

Finally, we observe that TIS's performance varies across different products, influenced by the characteristics of their social networks. 
Datasets like Spark Thinking and Intelligent Floor Scrubber have large, information-rich networks with 335k and 138k interactions, and follower variance exceeding 130, allowing TIS to effectively identify suitable influencers. 
In contrast, TIS performs worse with SUPOR Boosted Showerhead due to its smaller network (94k interactions) and lower interaction variance (98), making it challenging to differentiate influencers and gather insights.

\section{Analysis and Discussion}

\subsection{UT and CL Modeling}
The user timeline and content lifecycle modeling are illustrated in Figure \ref{fig:utr}.
From Figure \ref{fig:utr}(a), we observe that the GMM effectively simulates the activity pattern of User U, who is engaged on social media platforms around noon, at 2 PM, and from 8 PM to 11 PM, remaining inactive during other times. 
When simulating his behavior with the LLM, we take the agent offline during these inactive periods. 
Figure \ref{fig:utr}(b) displays the survival probabilities of 200 randomly sampled pieces of content (including posts and comments) within a one-day cycle. 
The figure reveals that most content becomes inactive within the first 10 minutes, some content gradually enters the death phase over time, while only a small fraction remains active for an extended period. 
Content with a long survival cycle is crucial for the interaction network, yet it represents a small quantity. 
By implementing content lifecycle modeling for the survival analysis, we effectively reduce the simulation burden on our system by limiting the influence of content likely to become inactive.

\subsection{Interaction Graph Expansion}
Figure \ref{fig:case} shows the expansion of the interaction network after four time cycles. The overall trend of the generated comments is generally positive, with examples like ``I'm here to show support!'' and ``I love it so much too!''. However, a minority express doubts and concerns, such as ``Is this cream really effective?'' and ``A bit pricey, any discounts available?''. This emotional distribution is influenced by Weibo's ecosystem, where influencers' fans are more inclined to leave positive comments, such as the real examples ``This dress is so beautiful!'' and ``I love their bath oil, it's a must-have for winter.''. 
These types of comments largely shape the simulated interactions. 

However, unlike real comments, our simulated ones lack daily greetings such as ``You'll get better and better'' and ``I've always been here'', and instead focus more on direct product inquiry. 
This is because our simulation framework is based on user profiles. 
Users inclined towards greetings tend to ignore advertisements and choose the ``ignore'' action. 
This explains the difference in the distribution of our simulated comments compared to the real advertising environment.

\begin{table}[htbp]
\small
  \centering
  \resizebox{\linewidth}{!}{
\begin{tabular}[htbp]{cccccc}
\toprule
Products & Influencers & Agents & $\text{Agents}_{T=0}$ & $V_{T=100}$ & $E_{T=100}$ \\
\midrule
Spark Thinking & 55    & 53,025 & 260   & 602 & 1,549 \\
Intelligent Floor Scrubber & 190   & 60,513 & 338   & 303 & 397 \\
Ruby Face Cream & 201   & 59,851 & 1,775  & 592 & 1,275 \\
SUPOR Boosted Showerhead & 81    & 38,433 & 190   & 451 & 1,030 \\
\bottomrule
\end{tabular}}
\caption{The statistics of the simulation scale. ``Influencers'' denotes the KOLs involved in the four products, ``Agents'' denotes the number of simulated agents, ``$\text{Agents}_{T=0}$'' denotes the number of agents in period $T=0$, and $<V_{T=100}, E_{T=100}>$ denote the scale of the interaction network formed after $T=100$ periods, where $V_{T=100}$ represents the vertex of the graph and $E_{T=100}$ represents the interaction edges.}
    \label{tab:simulate}%
\end{table}%

\subsection{Simulation Scale Analysis}
Table \ref{tab:simulate} shows the scale of the simulation for agents in various domains. 
We observe that the Intelligent Floor Scrubber has the largest number of agents, but after 100 periods, the extended interaction network is the smallest. 
This indicates that genuine comments in the technology domain have less positive sentiment compared to the parenting domain Spark Thinking. 
With equal scale, agents of Spark Thinking are more inclined to comment rather than ignore.

From the differences of $\text{Agents}_{T=0}$, we can see that at $T=0$, which is in the early morning, beauty domain users are more active on the internet. 
The least active is the home domain, as shown by the SUPOR Boosted Showerhead.

Looking at the interaction networks formed after 100 periods, the average number of interaction rounds for users has reached two, except for the Intelligent Floor Scrubber. 
In real scenarios, the number of interaction rounds per user for Ruby Face Cream and SUPOR Boosted Showerhead is also around two ~\cite{zhang2024sagraphlargescaletextrichsocial}, indicating that our simulated interaction scale is quite similar to real-life situations.

Although the overall agent scale is relatively large, 90\% of the agents remain in the self-awareness stage and do not take any actions related to the advertisement. 
At the same time, we have spread out the requests for simulating behaviors at various times through their time trajectories, significantly reducing the pressure on the 10\% that generate interactions and self-assessment. 
By keeping the request concurrency at 100, we complete the simulation operations for all influencers under each product within one hour.

\begin{figure}[tb]
    \centering
\includegraphics[width=0.75\linewidth]{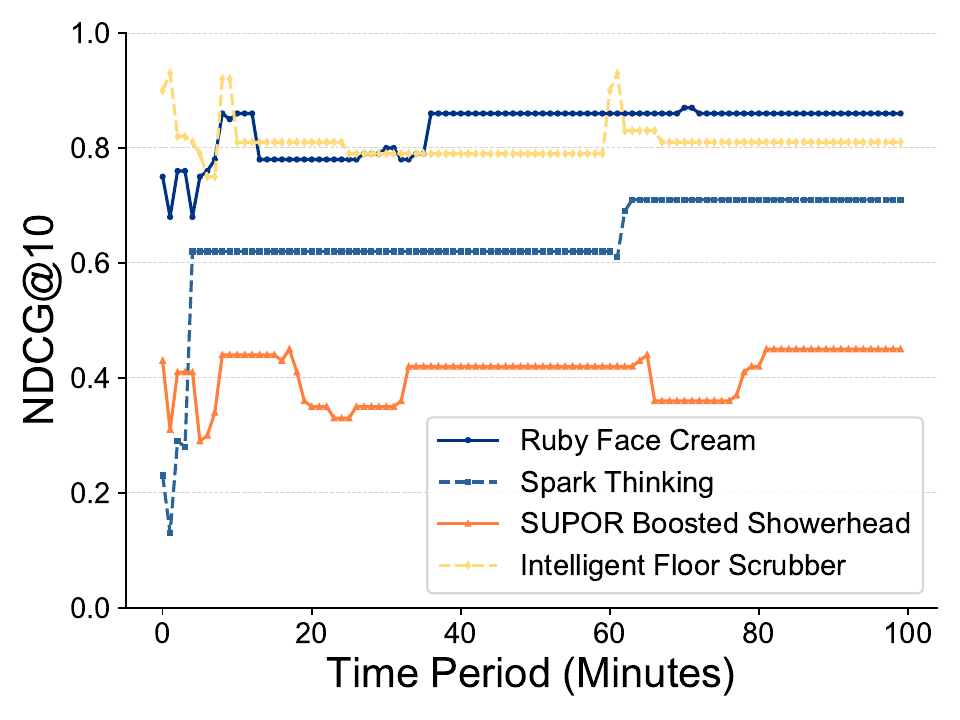} 
    \caption{The changes in the NDCG@10 metric over time in the influencer selection task for Advertising Campaigns, which illustrates the variation curves of four products on the SAGraph dataset. These curves represent the social networks across four different domains.}
    \label{fig:time}
\end{figure}

\subsection{Time Scale Analysis}
As time progresses, the performance of the promoters' recommendations is illustrated in Figure \ref{fig:time}. We set the simulation period to one minute, and after approximately ten simulation cycles, the performance of these recommendations reached a higher standard. This indicates that we have identified the most influential influencers relevant to recommending specific products within the field. 
As the simulation continues, the effectiveness of recommendations for products in different categories shows varying trends. For instance, recommendations for Spark Thinking steadily increase, while those for the SUPOR Boosted Showerhead remain at a relatively lower success rate. The Ruby Face Cream experiences short-term fluctuations, and the Intelligent Floor Scrubber exhibits the most significant variation. However, all products maintain relatively good recommendation effectiveness over an extended simulation period.
These fluctuations reflect the characteristics of user data in their respective domains. For example, Ruby Face Cream appeals to a diverse range of Key Opinion Leaders (KOLs) in the beauty sector. After a certain simulation period, as more KOLs emerge, our recommended rankings align more closely with expectations. The consistent rise of Spark Thinking can be attributed to the broader user base in parenting, which encompasses numerous KOLs, necessitating a longer period to identify the most influential leaders. 

\section{Conclusion}
In this study, we introduced the Time-aware Influencer Simulator
(TIS), an innovative LLM-based framework for influencer selection in advertising campaigns. 
Our framework leverages a large agent society over time that accurately mimics real-world advertising networks, including social relationships, content, and user behaviors, to simulate the nuanced dynamics of an influencer's interaction network by integrating a temporal trajectory simulation dimension. 
With a unique ranking metric, the TIS effectively identifies key influencers with the highest purchase inclination, aligning well with promoter preferences and offering rich insights into comments and attitudes. 
This research highlights the immense potential of LLMs in social recommendation and advertising, offering a valuable tool for promoters in decision-making processes.


\bibliographystyle{ACM-Reference-Format}
\bibliography{sample-base}


\begin{thebibliography}{47}


\ifx \showCODEN    \undefined \def \showCODEN     #1{\unskip}     \fi
\ifx \showDOI      \undefined \def \showDOI       #1{#1}\fi
\ifx \showISBNx    \undefined \def \showISBNx     #1{\unskip}     \fi
\ifx \showISBNxiii \undefined \def \showISBNxiii  #1{\unskip}     \fi
\ifx \showISSN     \undefined \def \showISSN      #1{\unskip}     \fi
\ifx \showLCCN     \undefined \def \showLCCN      #1{\unskip}     \fi
\ifx \shownote     \undefined \def \shownote      #1{#1}          \fi
\ifx \showarticletitle \undefined \def \showarticletitle #1{#1}   \fi
\ifx \showURL      \undefined \def \showURL       {\relax}        \fi
\providecommand\bibfield[2]{#2}
\providecommand\bibinfo[2]{#2}
\providecommand\natexlab[1]{#1}
\providecommand\showeprint[2][]{arXiv:#2}

\bibitem[Abu-Salih et~al\mbox{.}(2020)]%
        {abu2020time}
\bibfield{author}{\bibinfo{person}{Bilal Abu-Salih}, \bibinfo{person}{Kit~Yan Chan}, \bibinfo{person}{Omar Al-Kadi}, \bibinfo{person}{Marwan Al-Tawil}, \bibinfo{person}{Pornpit Wongthongtham}, \bibinfo{person}{Tomayess Issa}, \bibinfo{person}{Heba Saadeh}, \bibinfo{person}{Malak Al-Hassan}, \bibinfo{person}{Bushra Bremie}, {and} \bibinfo{person}{Abdulaziz Albahlal}.} \bibinfo{year}{2020}\natexlab{}.
\newblock \showarticletitle{Time-aware domain-based social influence prediction}.
\newblock \bibinfo{journal}{\emph{Journal of Big Data}}  \bibinfo{volume}{7} (\bibinfo{year}{2020}), \bibinfo{pages}{1--37}.
\newblock


\bibitem[Bender et~al\mbox{.}(2005)]%
        {bender2005generating}
\bibfield{author}{\bibinfo{person}{Ralf Bender}, \bibinfo{person}{Thomas Augustin}, {and} \bibinfo{person}{Maria Blettner}.} \bibinfo{year}{2005}\natexlab{}.
\newblock \showarticletitle{Generating survival times to simulate Cox proportional hazards models}.
\newblock \bibinfo{journal}{\emph{Statistics in medicine}} \bibinfo{volume}{24}, \bibinfo{number}{11} (\bibinfo{year}{2005}), \bibinfo{pages}{1713--1723}.
\newblock


\bibitem[Bohannon et~al\mbox{.}(2020)]%
        {bohannon2020autoregressive}
\bibfield{author}{\bibinfo{person}{Addison~W Bohannon}, \bibinfo{person}{Vernon~J Lawhern}, \bibinfo{person}{Nicholas~R Waytowich}, {and} \bibinfo{person}{Radu~V Balan}.} \bibinfo{year}{2020}\natexlab{}.
\newblock \showarticletitle{The autoregressive linear mixture model: A time-series model for an instantaneous mixture of network processes}.
\newblock \bibinfo{journal}{\emph{IEEE Transactions on Signal Processing}}  \bibinfo{volume}{68} (\bibinfo{year}{2020}), \bibinfo{pages}{4481--4496}.
\newblock


\bibitem[Chan(2022)]%
        {chan2022study}
\bibfield{author}{\bibinfo{person}{Fred Chan}.} \bibinfo{year}{2022}\natexlab{}.
\newblock \showarticletitle{A Study of Social Media Influencers and Impact on Consumer Buying Behaviour in the United Kingdom}.
\newblock \bibinfo{journal}{\emph{International Journal of Business \& Management Studies}} \bibinfo{volume}{3}, \bibinfo{number}{07} (\bibinfo{year}{2022}), \bibinfo{pages}{2694--1449}.
\newblock


\bibitem[Chen et~al\mbox{.}(2023a)]%
        {chen2023multi}
\bibfield{author}{\bibinfo{person}{Huaben Chen}, \bibinfo{person}{Wenkang Ji}, \bibinfo{person}{Lufeng Xu}, {and} \bibinfo{person}{Shiyu Zhao}.} \bibinfo{year}{2023}\natexlab{a}.
\newblock \showarticletitle{Multi-agent consensus seeking via large language models}.
\newblock \bibinfo{journal}{\emph{arXiv preprint arXiv:2310.20151}} (\bibinfo{year}{2023}).
\newblock


\bibitem[Chen et~al\mbox{.}(2023b)]%
        {chen2023topic}
\bibfield{author}{\bibinfo{person}{Xiuying Chen}, \bibinfo{person}{Mingzhe Li}, \bibinfo{person}{Shen Gao}, \bibinfo{person}{Xin Cheng}, \bibinfo{person}{Qiang Yang}, \bibinfo{person}{Qishen Zhang}, \bibinfo{person}{Xin Gao}, {and} \bibinfo{person}{Xiangliang Zhang}.} \bibinfo{year}{2023}\natexlab{b}.
\newblock \showarticletitle{A Topic-aware Summarization Framework with Different Modal Side Information}.
\newblock \bibinfo{journal}{\emph{SIGIR}} (\bibinfo{year}{2023}).
\newblock


\bibitem[Chen et~al\mbox{.}(2023c)]%
        {chen2023improving}
\bibfield{author}{\bibinfo{person}{Xiuying Chen}, \bibinfo{person}{Guodong Long}, \bibinfo{person}{Chongyang Tao}, \bibinfo{person}{Mingzhe Li}, \bibinfo{person}{Xin Gao}, \bibinfo{person}{Chengqi Zhang}, {and} \bibinfo{person}{Xiangliang Zhang}.} \bibinfo{year}{2023}\natexlab{c}.
\newblock \showarticletitle{Improving the Robustness of Summarization Systems with Dual Augmentation}.
\newblock \bibinfo{journal}{\emph{ACL}} (\bibinfo{year}{2023}).
\newblock


\bibitem[Dutta et~al\mbox{.}(2020)]%
        {dutta2020deep}
\bibfield{author}{\bibinfo{person}{Subhabrata Dutta}, \bibinfo{person}{Sarah Masud}, \bibinfo{person}{Soumen Chakrabarti}, {and} \bibinfo{person}{Tanmoy Chakraborty}.} \bibinfo{year}{2020}\natexlab{}.
\newblock \showarticletitle{Deep exogenous and endogenous influence combination for social chatter intensity prediction}. In \bibinfo{booktitle}{\emph{Proceedings of the 26th ACM SIGKDD International Conference on Knowledge Discovery \& Data Mining}}. \bibinfo{pages}{1999--2008}.
\newblock


\bibitem[Fabra et~al\mbox{.}(2020)]%
        {fabra2020log}
\bibfield{author}{\bibinfo{person}{Javier Fabra}, \bibinfo{person}{Pedro {\'A}lvarez}, {and} \bibinfo{person}{Joaqu{\'\i}n Ezpeleta}.} \bibinfo{year}{2020}\natexlab{}.
\newblock \showarticletitle{Log-based session profiling and online behavioral prediction in E--Commerce websites}.
\newblock \bibinfo{journal}{\emph{IEEE Access}}  \bibinfo{volume}{8} (\bibinfo{year}{2020}), \bibinfo{pages}{171834--171850}.
\newblock


\bibitem[Gao et~al\mbox{.}(2024)]%
        {gao2024large}
\bibfield{author}{\bibinfo{person}{Chen Gao}, \bibinfo{person}{Xiaochong Lan}, \bibinfo{person}{Nian Li}, \bibinfo{person}{Yuan Yuan}, \bibinfo{person}{Jingtao Ding}, \bibinfo{person}{Zhilun Zhou}, \bibinfo{person}{Fengli Xu}, {and} \bibinfo{person}{Yong Li}.} \bibinfo{year}{2024}\natexlab{}.
\newblock \showarticletitle{Large language models empowered agent-based modeling and simulation: A survey and perspectives}.
\newblock \bibinfo{journal}{\emph{Humanities and Social Sciences Communications}} \bibinfo{volume}{11}, \bibinfo{number}{1} (\bibinfo{year}{2024}), \bibinfo{pages}{1--24}.
\newblock


\bibitem[Gao et~al\mbox{.}(2023)]%
        {gao2023s}
\bibfield{author}{\bibinfo{person}{Chen Gao}, \bibinfo{person}{Xiaochong Lan}, \bibinfo{person}{Zhihong Lu}, \bibinfo{person}{Jinzhu Mao}, \bibinfo{person}{Jinghua Piao}, \bibinfo{person}{Huandong Wang}, \bibinfo{person}{Depeng Jin}, {and} \bibinfo{person}{Yong Li}.} \bibinfo{year}{2023}\natexlab{}.
\newblock \showarticletitle{S\textsuperscript{3}: Social-network Simulation System with Large Language Model-Empowered Agents}.
\newblock \bibinfo{journal}{\emph{arXiv preprint arXiv:2307.14984}} (\bibinfo{year}{2023}).
\newblock


\bibitem[Goyal et~al\mbox{.}(2011)]%
        {goyal2011celf++}
\bibfield{author}{\bibinfo{person}{Amit Goyal}, \bibinfo{person}{Wei Lu}, {and} \bibinfo{person}{Laks~VS Lakshmanan}.} \bibinfo{year}{2011}\natexlab{}.
\newblock \showarticletitle{Celf++ optimizing the greedy algorithm for influence maximization in social networks}. In \bibinfo{booktitle}{\emph{Proceedings of the 20th international conference companion on World wide web}}. \bibinfo{pages}{47--48}.
\newblock


\bibitem[Gu et~al\mbox{.}(2017)]%
        {gu2017co}
\bibfield{author}{\bibinfo{person}{Yupeng Gu}, \bibinfo{person}{Yizhou Sun}, {and} \bibinfo{person}{Jianxi Gao}.} \bibinfo{year}{2017}\natexlab{}.
\newblock \showarticletitle{The co-evolution model for social network evolving and opinion migration}. In \bibinfo{booktitle}{\emph{Proceedings of the 23rd ACM SIGKDD international conference on knowledge discovery and data mining}}. \bibinfo{pages}{175--184}.
\newblock


\bibitem[He et~al\mbox{.}(2013)]%
        {he2013improving}
\bibfield{author}{\bibinfo{person}{Jingrui He}, \bibinfo{person}{Wei Shen}, \bibinfo{person}{Phani Divakaruni}, \bibinfo{person}{Laura Wynter}, {and} \bibinfo{person}{Rick Lawrence}.} \bibinfo{year}{2013}\natexlab{}.
\newblock \showarticletitle{Improving traffic prediction with tweet semantics}. In \bibinfo{booktitle}{\emph{Twenty-Third International Joint Conference on Artificial Intelligence}}. Citeseer.
\newblock


\bibitem[Iribarren and Moro(2007)]%
        {iribarren2007information}
\bibfield{author}{\bibinfo{person}{Jos{\'e}~Luis Iribarren} {and} \bibinfo{person}{Esteban Moro}.} \bibinfo{year}{2007}\natexlab{}.
\newblock \showarticletitle{Information diffusion epidemics in social networks}.
\newblock \bibinfo{journal}{\emph{arXiv preprint arXiv:0706.0641}} (\bibinfo{year}{2007}).
\newblock


\bibitem[Jia et~al\mbox{.}(2022)]%
        {jia2022tt}
\bibfield{author}{\bibinfo{person}{Wei Jia}, \bibinfo{person}{Ruizhe Ma}, \bibinfo{person}{Li Yan}, \bibinfo{person}{Weinan Niu}, {and} \bibinfo{person}{Zongmin Ma}.} \bibinfo{year}{2022}\natexlab{}.
\newblock \showarticletitle{TT-graph: A new model for building social network graphs from texts with time series}.
\newblock \bibinfo{journal}{\emph{Expert Systems With Applications}}  \bibinfo{volume}{192} (\bibinfo{year}{2022}), \bibinfo{pages}{116405}.
\newblock


\bibitem[Kaiya et~al\mbox{.}(2023)]%
        {kaiya2023lyfe}
\bibfield{author}{\bibinfo{person}{Zhao Kaiya}, \bibinfo{person}{Michelangelo Naim}, \bibinfo{person}{Jovana Kondic}, \bibinfo{person}{Manuel Cortes}, \bibinfo{person}{Jiaxin Ge}, \bibinfo{person}{Shuying Luo}, \bibinfo{person}{Guangyu~Robert Yang}, {and} \bibinfo{person}{Andrew Ahn}.} \bibinfo{year}{2023}\natexlab{}.
\newblock \showarticletitle{Lyfe agents: Generative agents for low-cost real-time social interactions}.
\newblock \bibinfo{journal}{\emph{arXiv preprint arXiv:2310.02172}} (\bibinfo{year}{2023}).
\newblock


\bibitem[Kalliovirta et~al\mbox{.}(2015)]%
        {kalliovirta2015gaussian}
\bibfield{author}{\bibinfo{person}{Leena Kalliovirta}, \bibinfo{person}{Mika Meitz}, {and} \bibinfo{person}{Pentti Saikkonen}.} \bibinfo{year}{2015}\natexlab{}.
\newblock \showarticletitle{A Gaussian mixture autoregressive model for univariate time series}.
\newblock \bibinfo{journal}{\emph{Journal of Time Series Analysis}} \bibinfo{volume}{36}, \bibinfo{number}{2} (\bibinfo{year}{2015}), \bibinfo{pages}{247--266}.
\newblock


\bibitem[Kirk and Dianov(2015)]%
        {kirk2015towards}
\bibfield{author}{\bibinfo{person}{Nicholas~H Kirk} {and} \bibinfo{person}{Ilya Dianov}.} \bibinfo{year}{2015}\natexlab{}.
\newblock \showarticletitle{Towards predicting first daily departure times: a gaussian modeling approach for load shift forecasting}.
\newblock \bibinfo{journal}{\emph{arXiv preprint arXiv:1507.04502}} (\bibinfo{year}{2015}).
\newblock


\bibitem[Lahiri et~al\mbox{.}(2021)]%
        {lahiri2021todd}
\bibfield{author}{\bibinfo{person}{Aditya Lahiri}, \bibinfo{person}{Yash~Kumar Singhal}, {and} \bibinfo{person}{Adwitiya Sinha}.} \bibinfo{year}{2021}\natexlab{}.
\newblock \showarticletitle{TODD: Time-aware opinion dynamics diffusion model for online social networks}. In \bibinfo{booktitle}{\emph{Proceedings of International Conference on Artificial Intelligence and Applications: ICAIA 2020}}. Springer, \bibinfo{pages}{235--245}.
\newblock


\bibitem[Lampos et~al\mbox{.}(2014)]%
        {lampos2014predicting}
\bibfield{author}{\bibinfo{person}{Vasileios Lampos}, \bibinfo{person}{Nikolaos Aletras}, \bibinfo{person}{Daniel Preo{\c{t}}iuc-Pietro}, {and} \bibinfo{person}{Trevor Cohn}.} \bibinfo{year}{2014}\natexlab{}.
\newblock \showarticletitle{Predicting and characterising user impact on Twitter}. In \bibinfo{booktitle}{\emph{Proceedings of the 14th Conference of the European Chapter of the Association for Computational Linguistics}}. \bibinfo{pages}{405--413}.
\newblock


\bibitem[Lenger(2022)]%
        {lenger2022choose}
\bibfield{author}{\bibinfo{person}{Asl{\i}~Diyadin Lenger}.} \bibinfo{year}{2022}\natexlab{}.
\newblock \showarticletitle{How to choose the right influencer for a marketing strategy}.
\newblock \bibinfo{journal}{\emph{Applied Marketing Analytics}} \bibinfo{volume}{8}, \bibinfo{number}{1} (\bibinfo{year}{2022}), \bibinfo{pages}{89--104}.
\newblock


\bibitem[Leskovec et~al\mbox{.}(2007)]%
        {leskovec2007cost}
\bibfield{author}{\bibinfo{person}{Jure Leskovec}, \bibinfo{person}{Andreas Krause}, \bibinfo{person}{Carlos Guestrin}, \bibinfo{person}{Christos Faloutsos}, \bibinfo{person}{Jeanne VanBriesen}, {and} \bibinfo{person}{Natalie Glance}.} \bibinfo{year}{2007}\natexlab{}.
\newblock \showarticletitle{Cost-effective outbreak detection in networks}. In \bibinfo{booktitle}{\emph{Proceedings of the 13th ACM SIGKDD international conference on Knowledge discovery and data mining}}. \bibinfo{pages}{420--429}.
\newblock


\bibitem[Liu et~al\mbox{.}(2019)]%
        {liu2019anomaly}
\bibfield{author}{\bibinfo{person}{Jianwei Liu}, \bibinfo{person}{Hongwei Zhu}, \bibinfo{person}{Yongxia Liu}, \bibinfo{person}{Haobo Wu}, \bibinfo{person}{Yunsheng Lan}, {and} \bibinfo{person}{Xinyu Zhang}.} \bibinfo{year}{2019}\natexlab{}.
\newblock \showarticletitle{Anomaly detection for time series using temporal convolutional networks and Gaussian mixture model}. In \bibinfo{booktitle}{\emph{Journal of Physics: Conference Series}}, Vol.~\bibinfo{volume}{1187}. IOP Publishing, \bibinfo{pages}{042111}.
\newblock


\bibitem[Liu et~al\mbox{.}(2024a)]%
        {liu2024skepticism}
\bibfield{author}{\bibinfo{person}{Yuhan Liu}, \bibinfo{person}{Xiuying Chen}, \bibinfo{person}{Xiaoqing Zhang}, \bibinfo{person}{Xing Gao}, \bibinfo{person}{Ji Zhang}, {and} \bibinfo{person}{Rui Yan}.} \bibinfo{year}{2024}\natexlab{a}.
\newblock \showarticletitle{From Skepticism to Acceptance: Simulating the Attitude Dynamics Toward Fake News}.
\newblock \bibinfo{journal}{\emph{IJCAI}} (\bibinfo{year}{2024}).
\newblock


\bibitem[Liu et~al\mbox{.}(2024b)]%
        {liu2024tiny}
\bibfield{author}{\bibinfo{person}{Yuhan Liu}, \bibinfo{person}{Zirui Song}, \bibinfo{person}{Xiaoqing Zhang}, \bibinfo{person}{Xiuying Chen}, {and} \bibinfo{person}{Rui Yan}.} \bibinfo{year}{2024}\natexlab{b}.
\newblock \showarticletitle{From a Tiny Slip to a Giant Leap: An LLM-Based Simulation for Fake News Evolution}.
\newblock \bibinfo{journal}{\emph{arXiv preprint arXiv:2410.19064}} (\bibinfo{year}{2024}).
\newblock


\bibitem[Mallipeddi et~al\mbox{.}(2022)]%
        {mallipeddi2022framework}
\bibfield{author}{\bibinfo{person}{Rakesh~R Mallipeddi}, \bibinfo{person}{Subodha Kumar}, \bibinfo{person}{Chelliah Sriskandarajah}, {and} \bibinfo{person}{Yunxia Zhu}.} \bibinfo{year}{2022}\natexlab{}.
\newblock \showarticletitle{A framework for analyzing influencer marketing in social networks: selection and scheduling of influencers}.
\newblock \bibinfo{journal}{\emph{Management Science}} \bibinfo{volume}{68}, \bibinfo{number}{1} (\bibinfo{year}{2022}), \bibinfo{pages}{75--104}.
\newblock


\bibitem[Park et~al\mbox{.}(2023)]%
        {park2023generative}
\bibfield{author}{\bibinfo{person}{Joon~Sung Park}, \bibinfo{person}{Joseph O'Brien}, \bibinfo{person}{Carrie~Jun Cai}, \bibinfo{person}{Meredith~Ringel Morris}, \bibinfo{person}{Percy Liang}, {and} \bibinfo{person}{Michael~S Bernstein}.} \bibinfo{year}{2023}\natexlab{}.
\newblock \showarticletitle{Generative agents: Interactive simulacra of human behavior}. In \bibinfo{booktitle}{\emph{Proceedings of the 36th annual acm symposium on user interface software and technology}}. \bibinfo{pages}{1--22}.
\newblock


\bibitem[Park et~al\mbox{.}(2022)]%
        {park2022social}
\bibfield{author}{\bibinfo{person}{Joon~Sung Park}, \bibinfo{person}{Lindsay Popowski}, \bibinfo{person}{Carrie Cai}, \bibinfo{person}{Meredith~Ringel Morris}, \bibinfo{person}{Percy Liang}, {and} \bibinfo{person}{Michael~S Bernstein}.} \bibinfo{year}{2022}\natexlab{}.
\newblock \showarticletitle{Social simulacra: Creating populated prototypes for social computing systems}. In \bibinfo{booktitle}{\emph{Proceedings of the 35th Annual ACM Symposium on User Interface Software and Technology}}. \bibinfo{pages}{1--18}.
\newblock


\bibitem[Peng et~al\mbox{.}(2018)]%
        {peng2018influence}
\bibfield{author}{\bibinfo{person}{Sancheng Peng}, \bibinfo{person}{Yongmei Zhou}, \bibinfo{person}{Lihong Cao}, \bibinfo{person}{Shui Yu}, \bibinfo{person}{Jianwei Niu}, {and} \bibinfo{person}{Weijia Jia}.} \bibinfo{year}{2018}\natexlab{}.
\newblock \showarticletitle{Influence analysis in social networks: A survey}.
\newblock \bibinfo{journal}{\emph{Journal of Network and Computer Applications}}  \bibinfo{volume}{106} (\bibinfo{year}{2018}), \bibinfo{pages}{17--32}.
\newblock


\bibitem[Qiu et~al\mbox{.}(2018)]%
        {qiu2018deepinf}
\bibfield{author}{\bibinfo{person}{Jiezhong Qiu}, \bibinfo{person}{Jian Tang}, \bibinfo{person}{Hao Ma}, \bibinfo{person}{Yuxiao Dong}, \bibinfo{person}{Kuansan Wang}, {and} \bibinfo{person}{Jie Tang}.} \bibinfo{year}{2018}\natexlab{}.
\newblock \showarticletitle{Deepinf: Social influence prediction with deep learning}. In \bibinfo{booktitle}{\emph{Proceedings of the 24th ACM SIGKDD international conference on knowledge discovery \& data mining}}. \bibinfo{pages}{2110--2119}.
\newblock


\bibitem[Raghavan et~al\mbox{.}(2014)]%
        {raghavan2014modeling}
\bibfield{author}{\bibinfo{person}{Vasanthan Raghavan}, \bibinfo{person}{Greg Ver~Steeg}, \bibinfo{person}{Aram Galstyan}, {and} \bibinfo{person}{Alexander~G Tartakovsky}.} \bibinfo{year}{2014}\natexlab{}.
\newblock \showarticletitle{Modeling temporal activity patterns in dynamic social networks}.
\newblock \bibinfo{journal}{\emph{IEEE Transactions on Computational Social Systems}} \bibinfo{volume}{1}, \bibinfo{number}{1} (\bibinfo{year}{2014}), \bibinfo{pages}{89--107}.
\newblock


\bibitem[Reynolds et~al\mbox{.}(2009)]%
        {reynolds2009gaussian}
\bibfield{author}{\bibinfo{person}{Douglas~A Reynolds} {et~al\mbox{.}}} \bibinfo{year}{2009}\natexlab{}.
\newblock \showarticletitle{Gaussian mixture models.}
\newblock \bibinfo{journal}{\emph{Encyclopedia of biometrics}} \bibinfo{volume}{741}, \bibinfo{number}{659-663} (\bibinfo{year}{2009}).
\newblock


\bibitem[Rivadeneira et~al\mbox{.}(2021)]%
        {rivadeneira2021predicting}
\bibfield{author}{\bibinfo{person}{Luc{\'\i}a Rivadeneira}, \bibinfo{person}{Jian-Bo Yang}, {and} \bibinfo{person}{Manuel L{\'o}pez-Ib{\'a}{\~n}ez}.} \bibinfo{year}{2021}\natexlab{}.
\newblock \showarticletitle{Predicting tweet impact using a novel evidential reasoning prediction method}.
\newblock \bibinfo{journal}{\emph{Expert Systems with Applications}}  \bibinfo{volume}{169} (\bibinfo{year}{2021}), \bibinfo{pages}{114400}.
\newblock


\bibitem[Salamzadeh et~al\mbox{.}(2022)]%
        {salamzadeh2022grocery}
\bibfield{author}{\bibinfo{person}{Aidin Salamzadeh}, \bibinfo{person}{Pejman Ebrahimi}, \bibinfo{person}{Maryam Soleimani}, {and} \bibinfo{person}{Maria Fekete-Farkas}.} \bibinfo{year}{2022}\natexlab{}.
\newblock \showarticletitle{Grocery apps and consumer purchase behavior: application of Gaussian mixture model and multi-layer perceptron algorithm}.
\newblock \bibinfo{journal}{\emph{Journal of Risk and Financial Management}} \bibinfo{volume}{15}, \bibinfo{number}{10} (\bibinfo{year}{2022}), \bibinfo{pages}{424}.
\newblock


\bibitem[Sharma and Gupta(2022)]%
        {sharma2022role}
\bibfield{author}{\bibinfo{person}{Saurabh Sharma} {and} \bibinfo{person}{Vishal Gupta}.} \bibinfo{year}{2022}\natexlab{}.
\newblock \showarticletitle{Role of twitter user profile features in retweet prediction for big data streams}.
\newblock \bibinfo{journal}{\emph{Multimedia Tools and Applications}} \bibinfo{volume}{81}, \bibinfo{number}{19} (\bibinfo{year}{2022}), \bibinfo{pages}{27309--27338}.
\newblock


\bibitem[Singh et~al\mbox{.}(2020)]%
        {singh2020predicting}
\bibfield{author}{\bibinfo{person}{Nidhi Singh}, \bibinfo{person}{Anurag Singh}, {and} \bibinfo{person}{Rajesh Sharma}.} \bibinfo{year}{2020}\natexlab{}.
\newblock \showarticletitle{Predicting information cascade on twitter using random walk}.
\newblock \bibinfo{journal}{\emph{Procedia Computer Science}}  \bibinfo{volume}{173} (\bibinfo{year}{2020}), \bibinfo{pages}{201--209}.
\newblock


\bibitem[T{\"o}rnberg et~al\mbox{.}(2023)]%
        {tornberg2023simulating}
\bibfield{author}{\bibinfo{person}{Petter T{\"o}rnberg}, \bibinfo{person}{Diliara Valeeva}, \bibinfo{person}{Justus Uitermark}, {and} \bibinfo{person}{Christopher Bail}.} \bibinfo{year}{2023}\natexlab{}.
\newblock \showarticletitle{Simulating social media using large language models to evaluate alternative news feed algorithms}.
\newblock \bibinfo{journal}{\emph{arXiv preprint arXiv:2310.05984}} (\bibinfo{year}{2023}).
\newblock


\bibitem[Wall(1991)]%
        {wall1991predicting}
\bibfield{author}{\bibinfo{person}{David~W Wall}.} \bibinfo{year}{1991}\natexlab{}.
\newblock \showarticletitle{Predicting program behavior using real or estimated profiles}.
\newblock \bibinfo{journal}{\emph{ACM SIGPLAN Notices}} \bibinfo{volume}{26}, \bibinfo{number}{6} (\bibinfo{year}{1991}), \bibinfo{pages}{59--70}.
\newblock


\bibitem[Wang et~al\mbox{.}(2024)]%
        {wang2024decoding}
\bibfield{author}{\bibinfo{person}{Chenxi Wang}, \bibinfo{person}{Zongfang Liu}, \bibinfo{person}{Dequan Yang}, {and} \bibinfo{person}{Xiuying Chen}.} \bibinfo{year}{2024}\natexlab{}.
\newblock \showarticletitle{Decoding Echo Chambers: LLM-Powered Simulations Revealing Polarization in Social Networks}.
\newblock \bibinfo{journal}{\emph{arXiv preprint arXiv:2409.19338}} (\bibinfo{year}{2024}).
\newblock


\bibitem[Wang et~al\mbox{.}(2022)]%
        {wang2022influential}
\bibfield{author}{\bibinfo{person}{Feng Wang}, \bibinfo{person}{Xueting Zhang}, \bibinfo{person}{Man Chen}, \bibinfo{person}{Wei Zeng}, {and} \bibinfo{person}{Rong Cao}.} \bibinfo{year}{2022}\natexlab{}.
\newblock \showarticletitle{The influential paradox: Brand and deal content sharing by influencers in friendship networks}.
\newblock \bibinfo{journal}{\emph{Journal of Business Research}}  \bibinfo{volume}{150} (\bibinfo{year}{2022}), \bibinfo{pages}{503--514}.
\newblock


\bibitem[Yan et~al\mbox{.}(2019)]%
        {yan2019minimizing}
\bibfield{author}{\bibinfo{person}{Ruidong Yan}, \bibinfo{person}{Deying Li}, \bibinfo{person}{Weili Wu}, \bibinfo{person}{Ding-Zhu Du}, {and} \bibinfo{person}{Yongcai Wang}.} \bibinfo{year}{2019}\natexlab{}.
\newblock \showarticletitle{Minimizing influence of rumors by blockers on social networks: algorithms and analysis}.
\newblock \bibinfo{journal}{\emph{IEEE Transactions on Network Science and Engineering}} \bibinfo{volume}{7}, \bibinfo{number}{3} (\bibinfo{year}{2019}), \bibinfo{pages}{1067--1078}.
\newblock


\bibitem[Yang and Leskovec(2010)]%
        {yang2010modeling}
\bibfield{author}{\bibinfo{person}{Jaewon Yang} {and} \bibinfo{person}{Jure Leskovec}.} \bibinfo{year}{2010}\natexlab{}.
\newblock \showarticletitle{Modeling information diffusion in implicit networks}. In \bibinfo{booktitle}{\emph{2010 IEEE international conference on data mining}}. IEEE, \bibinfo{pages}{599--608}.
\newblock


\bibitem[Zhang et~al\mbox{.}(2015)]%
        {zhang2015influenced}
\bibfield{author}{\bibinfo{person}{Jing Zhang}, \bibinfo{person}{Jie Tang}, \bibinfo{person}{Juanzi Li}, \bibinfo{person}{Yang Liu}, {and} \bibinfo{person}{Chunxiao Xing}.} \bibinfo{year}{2015}\natexlab{}.
\newblock \showarticletitle{Who influenced you? predicting retweet via social influence locality}.
\newblock \bibinfo{journal}{\emph{ACM Transactions on Knowledge Discovery from Data (TKDD)}} \bibinfo{volume}{9}, \bibinfo{number}{3} (\bibinfo{year}{2015}), \bibinfo{pages}{1--26}.
\newblock


\bibitem[Zhang et~al\mbox{.}(2024)]%
        {zhang2024sagraphlargescaletextrichsocial}
\bibfield{author}{\bibinfo{person}{Xiaoqing Zhang}, \bibinfo{person}{Xiuying Chen}, \bibinfo{person}{Yuhan Liu}, \bibinfo{person}{Jianzhou Wang}, \bibinfo{person}{Zhenxing Hu}, {and} \bibinfo{person}{Rui Yan}.} \bibinfo{year}{2024}\natexlab{}.
\newblock \bibinfo{title}{SAGraph: A Large-scale Text-Rich Social Graph Dataset for Advertising Campaigns}.
\newblock
\newblock
\showeprint[arxiv]{2403.15105}~[cs.SI]
\urldef\tempurl%
\url{https://arxiv.org/abs/2403.15105}
\showURL{%
\tempurl}


\bibitem[Zhou and Zhang(2023)]%
        {zhou2023opinion}
\bibfield{author}{\bibinfo{person}{Xiaotian Zhou} {and} \bibinfo{person}{Zhongzhi Zhang}.} \bibinfo{year}{2023}\natexlab{}.
\newblock \showarticletitle{Opinion Maximization in Social Networks via Leader Selection}. In \bibinfo{booktitle}{\emph{Proceedings of the ACM Web Conference 2023}}. \bibinfo{pages}{133--142}.
\newblock


\bibitem[Zhu et~al\mbox{.}(2021)]%
        {zhu2021minimizing}
\bibfield{author}{\bibinfo{person}{Liwang Zhu}, \bibinfo{person}{Qi Bao}, {and} \bibinfo{person}{Zhongzhi Zhang}.} \bibinfo{year}{2021}\natexlab{}.
\newblock \showarticletitle{Minimizing polarization and disagreement in social networks via link recommendation}.
\newblock \bibinfo{journal}{\emph{Advances in Neural Information Processing Systems}}  \bibinfo{volume}{34} (\bibinfo{year}{2021}), \bibinfo{pages}{2072--2084}.
\newblock


\end{thebibliography}










\end{document}